\pdfoutput=1
\documentclass[prx,twocolumn,superscriptaddress,nofootinbib,10pt]{revtex4-2}
\usepackage[caption=false]{subfig}
\usepackage{amsmath,amsfonts,amsthm,mathrsfs,dsfont,amssymb,graphicx,xspace,mathtools,array}
\usepackage[T1]{fontenc}
\usepackage{tikz}
\usepackage{textcomp}
\usepackage{enumitem}
\usepackage[scaled=0.9]{helvet}

\definecolor{linkblue}{HTML}{001487}
\usepackage[colorlinks=true,allcolors=linkblue]{hyperref}

\usepackage[nameinlink,capitalize]{cleveref}
\crefname{enumi}{Step}{Steps}
\setlist[enumerate]{label=\emph{\arabic*}., ref=\arabic*, leftmargin=12pt, topsep=2pt, itemsep=0pt}

\newtheorem{theorem}{Theorem}
\newtheorem*{theorem*}{Theorem}

\theoremstyle{remark}

\theoremstyle{definition}

\newtheorem{protocol}{Protocol}

\newcommand\numberthis{\addtocounter{equation}{1}\tag{\theequation}}

\newcommand{\ket}[1]{|#1\rangle}
\newcommand{\bra}[1]{\langle#1|}
\newcommand{\proj}[1]{\ket{#1}\!\bra{#1}}
\DeclarePairedDelimiterX\braket[2]{\langle}{\rangle}{#1 \delimsize\vert #2}

\newcommand{\pr}[1]{{\rm Pr}\!\left[ #1 \right]}

\newcommand{\1}{\mathds{1}}
\newcommand{\ot}{\ensuremath{\otimes}}

\newcommand{\C}{\ensuremath{\mathbb{C}}}

\newcommand{\mf}{\ensuremath{\mathcal{F}}}
\newcommand{\mg}{\ensuremath{\mathcal{G}}}
\let\H\relax
\newcommand{\H}{\ensuremath{\mathcal{H}}}

\newcommand{\mn}{\ensuremath{\mathcal{N}}}

\newcommand{\mx}{\ensuremath{\mathcal{X}}}
\newcommand{\my}{\ensuremath{\mathcal{Y}}}

\let\eps\varepsilon

\newcommand{\kg}{\ensuremath{\mathcal{K_G}}}
\newcommand{\kf}{\ensuremath{\mathcal{K_F}}}

\DeclareMathOperator{\negl}{negl}
\newcommand{\bits}{\ensuremath{\{0, 1\}}}

\newcommand{\ct}{\textsc{ct}}
\newcommand{\rt}{\textsc{rt}}

\usetikzlibrary{quantikz}

\newcommand{\changed}[1]{#1}

\begin{document}

\title{Device-independent quantum key distribution from computational assumptions}

\author{Tony Metger}
\email{tmetger@ethz.ch}
\affiliation{Institute for Theoretical Physics, ETH Zurich, 8092 Zurich, Switzerland}

\author{Yfke Dulek}
\affiliation{QuSoft, University of Amsterdam, the Netherlands}

\author{Andrea Coladangelo}
\affiliation{EECS Department, University of California Berkeley, USA}

\author{Rotem Arnon-Friedman}
\affiliation{Department of Physics of Complex Systems, Weizmann Institute of Science, Israel}

\begin{abstract}

In device-independent quantum key distribution (DIQKD), an adversary prepares a device consisting of two components, distributed to Alice and Bob, who use the device to generate a secure key.
The security of existing DIQKD schemes holds under the assumption that the two components of the device cannot communicate with one another during the protocol execution.
This is called the \emph{no-communication assumption} in DIQKD.
Here, we show how to replace this assumption, which can be hard to enforce in practice, by a standard \emph{computational assumption} from post-quantum cryptography:
we give a protocol that produces secure keys even when the components of an adversarial device can exchange arbitrary quantum communication, assuming the device is computationally bounded.
Importantly, the computational assumption only needs to hold during the protocol execution---the keys generated at the end of the protocol are \emph{information-theoretically secure} as in standard DIQKD protocols.
\end{abstract}

\maketitle

\section{Introduction} \label{sec:intro}
The security of classical public-key cryptography is based on the assumption that an adversary cannot solve a specific computational problem, e.g.~a lattice problem~\cite{peikert2016decade}.
A message encrypted with classical public-key cryptography only remains secret as long as this computational assumption holds.
If a faster algorithm or more powerful hardware allows the adversary to break the computational assumption in the future, all past communication is at risk.
In contrast, quantum key distribution (QKD) protocols generate keys that are \emph{information-theoretically secure}, i.e., secure even against an all-powerful adversary, and are not compromised by advances in algorithms or hardware.
The security of QKD protocols is based on certain assumptions (which depend on the specific protocol being considered) that need to hold \emph{during} the execution of the protocol---violating the assumptions afterwards does \emph{not} compromise the security of the key. 
This is known as \emph{everlasting security}~\cite{unruh2018everlasting}.

Early QKD protocols, such as the BB84 protocol~\cite{bennett1984proceedings}, relied on the assumption that the quantum device used to generate the key is implemented as intended. 
Any deviation from the implementation analysed in the security proof can potentially lead to a security breach~\cite{fung2007phase,lydersen2010hacking,weier2011quantum,gerhardt2011full}.
Device-independent QKD (DIQKD) protocols~\cite{ekert1991quantum, mayers1998quantum} address this problem: instead of making assumptions about the inner workings of the device, the device is being \emph{tested} as part of the protocol, as explained below.
This form of security is considered the ``gold-standard'' of quantum cryptography~\cite{ekert2014ultimate}. 
In particular, it allows for security statements that hold even when the manufacturer of the quantum device is incompetent or malicious.\footnote{A malicious manufacturer is one that produces a devices with the intention of gaining information about the supposedly secret key generated by the device.}

\begin{figure*}[ht!]

\subfloat[]{
\includegraphics[width=0.4\textwidth]{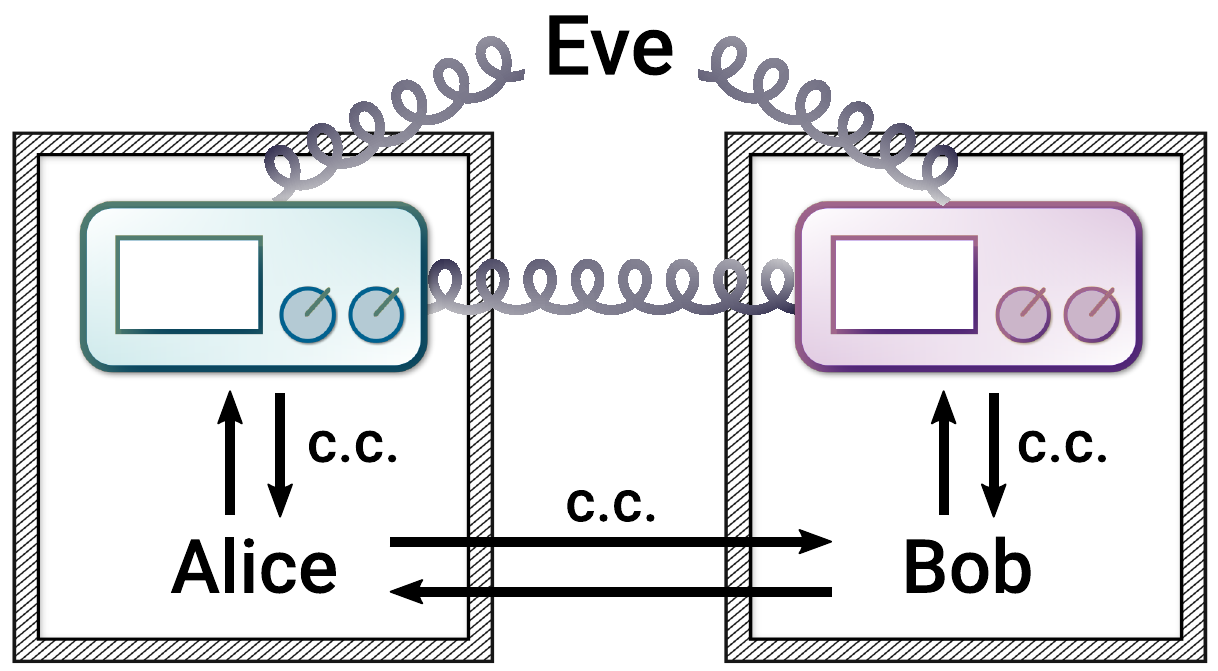}
\label{fig:setting_trad}
}
\qquad\qquad
\subfloat[]{
\includegraphics[width=0.4\textwidth]{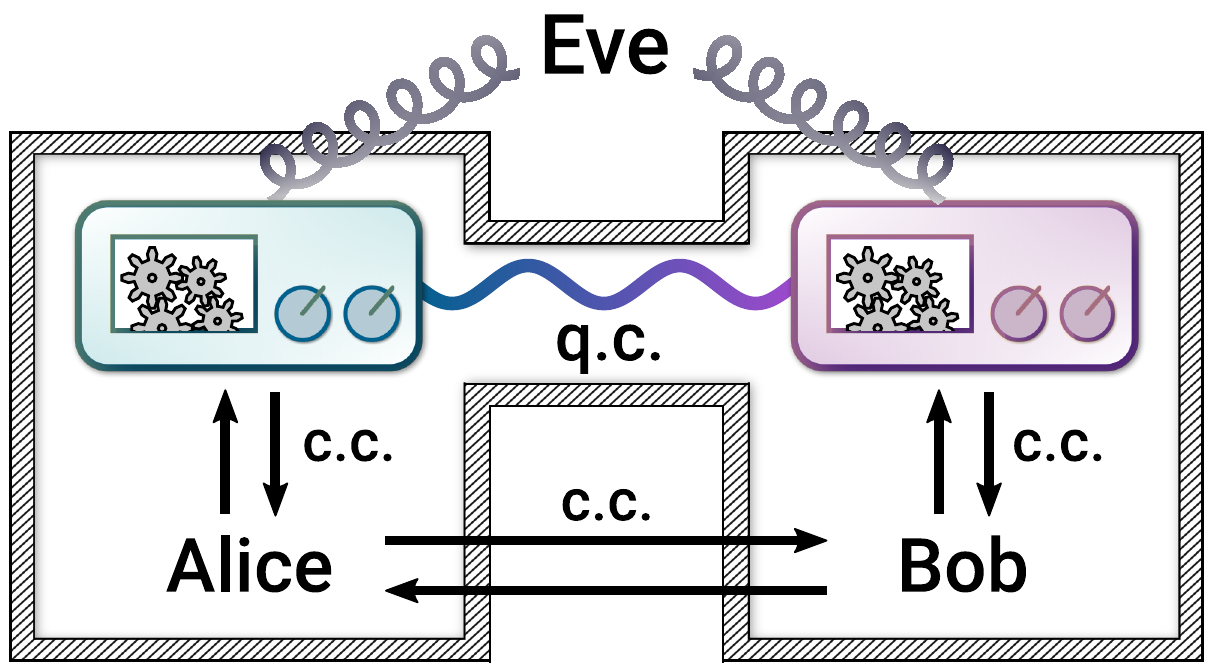}
\label{fig:setting_comp}
}

\caption{
{\bf (a) Traditional DIQKD setting.} 
Alice and Bob each receive a component of a device prepared by the adversary Eve. 
The device may use \emph{pre-shared entanglement} between its two components and can also be entangled with Eve (curly lines in the figure).
The components cannot signal to each other or to Eve during the protocol execution. 
Alice and Bob interact classically with the device and  communicate classically over a public authenticated  channel (straight lines labeled by c.c.\@ in the figure).
{\bf (b) Computational DIQKD setting.} The setting differs from the one in (a)  in two respects: 
(i) The two components are connected by a \emph{quantum channel} (labelled q.c.\@ in the figure); the channel is part of the device prepared by Eve, but, as the rest of the device, cannot be accessed by Eve after giving the device to Alice and Bob. 
(ii) It is assumed that the device is computationally bounded (denoted by gears in the figure) and cannot break post-quantum cryptographic problems \emph{during} the execution of the protocol, while Eve remains computationally unbounded as in (a).
}
\end{figure*}

The standard setting for DIQKD is shown in \cref{fig:setting_trad}.
Alice and Bob each hold a component of a device prepared by a potentially malicious party called Eve. 
Alice's and Bob's components of the device and the adversary Eve share some quantum state, and Alice's and Bob's components perform some quantum measurement on their respective part of the state.
The shared quantum state and the measurements  used by the components of the device are unknown to Alice and Bob.
Hence, the device is said to be uncharacterised.
Alice and Bob can only observe the classical \emph{input-output correlations} of their device: they supply classical inputs to the device (e.g., by pressing keys on a keyboard connected to the device) and receive classical outputs (e.g., by reading information displayed on its screen).

The security proofs of DIQKD protocols rely on the fact that a \emph{violation of a Bell inequality}~\cite{Bell1964, clauser1969proposed} can only be achieved by measuring an entangled quantum state.
Hence, if Alice and Bob observe that the classical input-output correlations of the device violate a Bell inequality, they can conclude, under certain conditions, that the two components must have been entangled.
This can then be used to certify the randomness, or entropy, of the output bits produced by measuring the entangled state. This certified entropy, in turn, acts as the basis for proving the security of the protocol~\cite{reichardt2013classical,vazirani2014fully,miller2014robust,arnon2018practical}.

For a security proof based on Bell inequalities to be valid, certain conditions must hold (or be assumed to hold). 
A potential violation of one of these conditions is called a \emph{loophole}.
An open loophole translates directly to a security breach in any DI cryptographic protocol~\cite{pironio2009device}. 
A fundamental and experimentally challenging loophole is the so-called \emph{communication loophole}---even a classical device can violate a Bell inequality when Alice's and Bob's  components can communicate. 
Therefore, to conclude that the correlations produced by the device must have arisen from measuring an entangled state, we must assume that the two components of the device cannot communicate \changed{during one round of the protocol}.  

There are two ways to experimentally enforce non-communication between the two components of the device.
The first is to make sure that the interaction with the device takes less time than a light signal needs to travel from one component to the other. Since special relativity forbids any signal to travel faster than light, this closes the communication loophole.
With technological limitations on the speed of, e.g., the production of entanglement or the usage of a random number generator in experiments, this approach requires a separation between Alice's and Bob's components on the order of kilometres, leading to additional experimental difficulties and constraints on where the protocol can be used (see, e.g.,~\cite{hensen2015experimental}).

A different approach is to physically shield the components of the device so that they cannot communicate with each other.
\changed{Without any communication at any point in the protocol}, \emph{all} the entangled particles required for a device to succeed in the protocol (at the very least $\sim\!\!10^7$ pairs) would need to be distributed \emph{prior} to the execution of the protocol and stored in Alice's and Bob's components.
Given the difficulties associated with storing quantum states, this is usually impractical.
Therefore, in typical implementations of QKD, the two components of the device are connected by a quantum channel, so that entanglement can be distributed ``on the fly'': one component creates an EPR pair and sends one qubit of the pair to the other component.
\changed{This makes shielding the components more difficult and the protocol execution potentially time-consuming as one has to be able to ``un-shield'' the components between rounds of the protocol to allow for entanglement distribution before ``re-shielding'' them for the  next round of the protocol.}

Given the difficulty of perfectly shielding components of a device from one another, recent works have aimed at formulating Bell inequalities that tolerate some limited amount of communication between the two components of the device~\cite{silman2013device, tavakoli2019informationally, tavakoli2020characterising}. 
For a key distribution scheme based on such a Bell inequality to be secure, one needs to assume an \emph{a priori} bound on the amount of communication; this bound cannot be verified during the protocol.
Hence, these works allow a weakening of the no-communication assumption, replacing the requirement of no communication by a requirement of limited communication.
Gaining confidence as to whether the weakened no-communication assumption, i.e., the bound on the amount of communication, holds must be done in a device-\emph{dependent} way and under various other assumptions. 

In this work, we are interested in whether the no-communication assumption is a necessary one for DIQKD, or whether DIQKD can also be based on the ``quantumness'' of the devices alone.
To study DIQKD without the no-communication assumption, we consider the setting in \cref{fig:setting_comp}. 
In this setting, the two components of the (untrusted) device are connected by a  quantum channel, modelling the channel used for on-the-fly entanglement distribution.
It is necessary to assume that Eve cannot access information sent via this channel, as otherwise this could be used by the device to signal to Eve.
We will additionally require that a protocol for the setting of \cref{fig:setting_comp} has an honest implementation that can also be executed in the setting of \cref{fig:setting_trad}, i.e., the honest implementation only requires pre-shared EPR pairs and local operations.
In other words, we include attacks that use the additional channel in our soundness analysis, while restricting to protocols that do not use it in our completeness proof.
We discuss these assumptions and requirements further in \cref{sec:discussion}.

Our main result is a DIQKD protocol (\cref{protocol:comp_qkd} described below) to generate \emph{information-theoretically secure keys} in the setting of \cref{fig:setting_comp}, assuming that the device in the protocol is \emph{computationally bounded}: we assume that the device cannot solve the Learning with Errors (LWE) problem,\footnote{Roughly speaking, the LWE problem corresponds to solving a noisy linear equation: given a matrix $A$ and a vector $b$ such that $A x + e = b$, where $e$ is a sufficiently short noise vector so that the solution~$x$ is unique, one needs to find $x$. For $e=0$, this is easily  solved by Gaussian elimination, but for a suitably sampled non-zero $e$, no efficient classical nor quantum algorithm is known.} a standard computational assumption in post-quantum cryptography~\cite{lwe, peikert2016decade}
More specifically, we assume that the device is computationally bounded and that the probability of any computationally bounded device to solve the LWE problem is negligible in the security parameter $\lambda$.\footnote{A negligible function is one that decays faster than any inverse polynomial. A security parameter quantifies how hard an instance of a cryptographic problem is. As a simple example, consider factoring: here, the security parameter could be the number of bits of the composite number that needs to be factored.}
This is called the \emph{LWE assumption}.

Crucially, unlike in classical public-key cryptography, this computational assumption can be leveraged to generate an information-theoretically secure key:
our DIQKD protocol (\cref{protocol:comp_qkd} below) achieves the same everlasting security as existing DIQKD protocols.
The security of our protocol relies on the fact that, much like the no-communication assumption and Bell inequalities in typical DIQKD protocols, the computational assumption in our setting gives Alice and Bob a way to test the device and certify that it uses entangled quantum states.
%Indeed, an \emph{honest} device can succeed in our protocol with the same non-local resources as those required for entanglement-based QKD protocols, i.e., EPR pairs.
%In other words, while we take into account that an arbitrary device might use quantum communication between its two components to cheat, the honest device is able to succeed using only EPR pairs (pre-shared or distributed on the fly) and does not need to apply non-local quantum operations.

When studying QKD protocols, the main quantity of interest is the \emph{key rate} of the protocol, namely, the length of the produced key divided by the number of rounds of the protocol. For simplicity, we consider the asymptotic key rate, which we denote by $K$, that describes the idealised case where one executes infinitely many rounds of the protocol and the device used in the protocol behaves independently and identically in each round (called the \emph{IID assumption}). The extension of our result to the setting of finitely many and possibly correlated repetitions of the protocol is briefly discussed in \cref{sec:discussion}.

Our main theorem sets a lower bound on the key rate of our DIQKD protocol. 
It involves two security parameters, $\eps$ and $\lambda$.
Roughly speaking, $\eps$ is the maximum probability with which a device is allowed to fail in one round of the DIQKD protocol (e.g. due to noise).
The higher the allowed value of $\eps$, the lower the key rate of the protocol will be.
The parameter $\lambda$ is a security parameter for the LWE problem.
The LWE assumption ensures that one can make the probability that the computationally bounded device solves the LWE problem during the protocol execution arbitrarily small, while still allowing an honest computationally bounded device to succeed in the protocol with probability close to 1.

\begin{theorem} \label{thm:security_iid}
Consider the setting of \cref{fig:setting_comp} and make the LWE assumption.
Suppose Alice and Bob execute the key distribution protocol, \cref{protocol:comp_qkd}, with threshold~$\eps$ and security parameter~$\lambda$.
Assume the device behaves independently and identically in each round of the protocol.
If the device leads the protocol to abort with probability approaching 0 as the number of rounds $n \to \infty$, then the key rate $K$ is at least
\begin{equation}
    K \geq \frac{1}{128} - O(\eps^c \log \eps) - \negl(\lambda)
\end{equation}
for a small constant $c$.
\end{theorem}

In the theorem, the notation $O(\eps^c \log \eps)$ means that there exists a constant $C$ such that this term is bounded from above in absolute value by $C \, \eps^c \log \eps$ for sufficiently small $\eps$. The term $\negl(\lambda)$ denotes a negligible function in $\lambda$, i.e., a function that decays faster than any inverse polynomial in $\lambda$.

Some remarks are in order.
Firstly, our DIQKD protocol (\cref{protocol:comp_qkd}) allows an honest device to succeed with probability negligibly (in $\lambda$) close to 1 using only EPR pairs, pre-shared or distributed on the fly, and local quantum operations.
Hence, for an honest device with pre-shared entanglement, the quantum channel between the two components of the device shown in \cref{fig:setting_comp} is not necessary, but a dishonest device may use it. 

Secondly, the constant $\frac{1}{128}$ is a consequence of fixing certain parameters in the protocol for simplicity. For practical implementations, these parameters could be optimized.

Thirdly, as mentioned before, while \cref{thm:security_iid} makes use of a computational assumption, encrypting a message with the resulting key differs fundamentally from classical public-key encryption.
The latter type of encrypted message can be intercepted and stored, with the purpose of decrypting it years later, once it becomes technologically feasible to break the computational assumption.
In contrast, the key rate $K$ in \cref{thm:security_iid} refers to an \emph{information-theoretically} secure key: unless the key generation device has enough computational power to break the computational assumption  \emph{during} the execution of the protocol, the encrypted message is guaranteed to be information-theoretically secure.
For a practical implementation, this means that unlike instances of computational problems in classical cryptography, which must be chosen large enough so that they cannot be solved for years to come, our protocol only requires comparatively small instances that are just large enough so that the device cannot solve them in the short time it takes to execute the protocol.

\begin{figure*}[ht!]

\subfloat[]{
\includegraphics[width=0.34\textwidth]{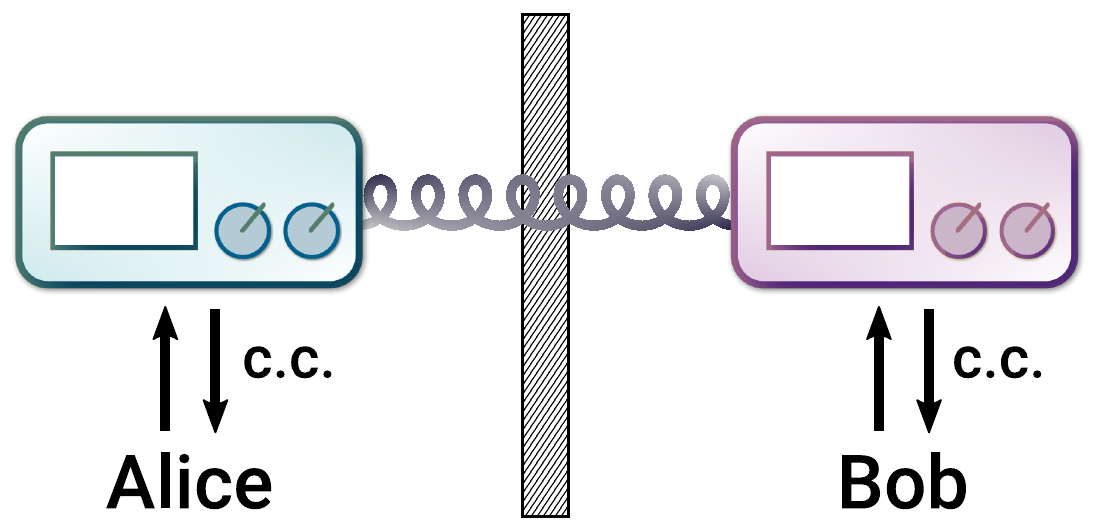}
\label{fig:self_testing_trad}
}
\qquad\qquad\quad
\subfloat[]{
\includegraphics[width=0.34\textwidth]{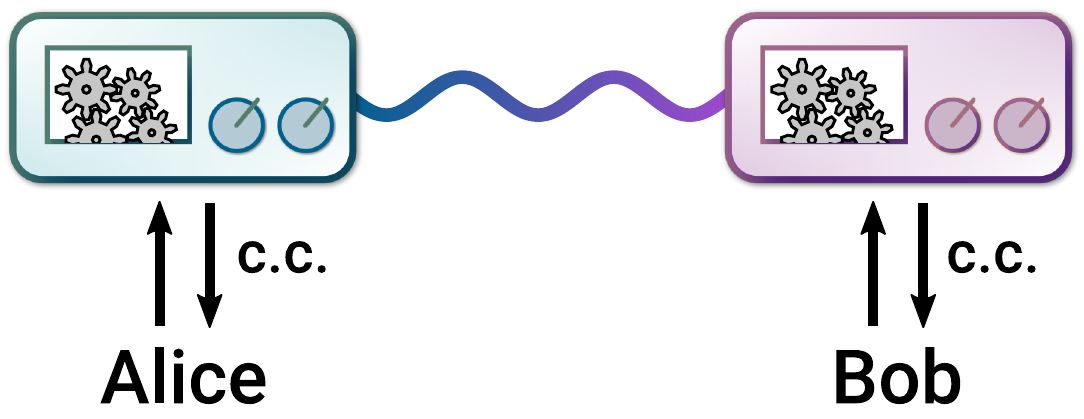}
\label{fig:self_testing_comp}
}

\caption{
{\bf (a) Traditional self-testing setting.} 
Alice and Bob each classically interact with a component of a quantum device. The two components of the device may be entangled, but cannot communicate.
{\bf (b) Computational self-testing setting.} Alice and Bob each classically interact with a component of a computationally bounded quantum device. The two components are connected by a quantum channel. This means that the device can implement any global operation on both of its components. Hence, the separation of the device into two components is arbitrary and we can equivalently treat it as a single device without internal structure.
}

\end{figure*}

\section{Computational self-testing \label{sec:comp_self_test}} 
We briefly review a recent protocol for self-testing~\cite{comp_self_testing}, a fundamental primitive in device-independent quantum information processing, of a single quantum device under computational assumptions. 
This will form the basis for our DIQKD protocol in \cref{sec:comp_qkd}. 

Self-testing \cite{sw87, pr92, mayers_yao, scarani-singlet, Coladangelo2017, supic_review} is a method in device-independent quantum information processing that certifies, only from classical input-output statistics of a quantum device, that a certain state and measurements must have been used to generate the device's output.
The setting for self-testing is the standard one for Bell experiments, pictured in \cref{fig:self_testing_trad}. 
Alice and Bob each receive a component of a quantum device.
The two components may be entangled, but cannot communicate with each other. 
Alice and Bob play a ``game'' with their device: they send (classical) \emph{questions} to the device, and the device returns (classical) \emph{answers}.
We say that the device has \emph{won the game} if it answers correctly according to some pre-defined \emph{winning condition}.

Let us denote the maximal winning probability for a quantum device by $\omega^*$.
A typical self-testing statement is as follows:
\emph{Assume the two components of the device cannot communicate and the device wins with probability close to $\omega^*$.
Then, up to local changes of basis in each component of the device and a small difference in trace distance, the device must have used a specific bipartite quantum state $\ket{\phi_{\rm ref}}_{AB}$, and specific measurements~$M_A$ for Alice's component and~$N_B$ for Bob's component.}\footnote{The distance between measurements needs to be measured in a special ``state-dependent distance''; see \cite[Section 4.4]{neexp} for a detailed exposition.}
For example, in the CHSH game, the \emph{reference state} $\ket{\phi_{\rm ref}}_{AB}$ is an EPR pair, the measurements~$M_A$ on Alice's side are computational or Hadamard basis measurements (depending on Alice's question), and the measurements~$N_B$ on Bob's side are similar (but rotated).

As explained earlier, the non-communication assumption is difficult to enforce in some experimental settings.
Motivated by this difficulty, a self-testing protocol that replaces the non-communication assumption by the assumption that the device is computationally bounded was introduced in \cite{comp_self_testing}, building on techniques from \cite{randomness, verification, rsp}.
This setting, shown in \cref{fig:self_testing_comp}, inspired our QKD setting in \cref{fig:setting_comp}.
The protocol from \cite{comp_self_testing} is described as \cref{protocol:comp_self_testing} below.

Note that this setting allows arbitrary quantum communication between the two components of the device, thereby opening the possibility for the device to perform any non-local (with respect to the two components) gate.
This setting is thus mathematically equivalent to the setting of a single device without any spatially separated components or other internal structure.
The protocol in~\cite{comp_self_testing}, which we build on, is presented in this ``single-device'' setting.

Before describing the protocol from \cite{comp_self_testing} in more detail, let us connect its result to standard self-testing statements of the form above.
The protocol from \cite{comp_self_testing} has multiple rounds of interaction between Alice, Bob, and the device.
For the purpose of self-testing, we are interested in the last round of interaction: here, Alice and Bob  send an input, a \emph{question}, $(x,y)$ to the device and receive an output, an \emph{answer}, $(a,b)$.
We can model the behaviour of the device in this last round by a quantum state $\sigma$ and measurements $\{P_{x, y}^{(a, b)}\}_{a, b}$, meaning that when the device receives questions $(x, y)$, it measures $\{P_{x, y}^{(a, b)}\}_{a, b}$ on $\sigma$ to obtain answers $(a, b)$.
The goal of \cref{protocol:comp_self_testing} is to ensure that the device's state $\sigma$ is a Bell state (i.e., the reference state $\phi_{\rm ref}$ for the protocol is a Bell state), and that the devices measurements $P_{x, y}^{(a, b)}$ are specific product measurements $(M_x^a)_A \ot (N_y^b)_B$, up to a change of basis and a small error. 

We now describe the self-testing protocol from \cite{comp_self_testing}.
The protocol makes use of a key and a trapdoor.
The key should be thought of as a piece of public information that specifies a particular instance of a cryptographic problem.
The trapdoor is a piece of private information with which the cryptographic problem can be solved efficiently.
Alice and Bob use such private trapdoors to be able to efficiently evaluate whether the device, which has no access to the trapdoor and is assumed to be unable to solve the cryptographic problem, has succeeded in the protocol or not. 
We also describe the behaviour of an \emph{honest} device, i.e., a device that behaves in the way Alice and Bob would like it to, omitting some details for the sake of brevity.
A more detailed description of the honest strategy for a modified version of this protocol, \cref{protocol:self_testing_teleportation} below, can be found in \cref{app:honest_behaviour}.

\begin{protocol}[Self-testing protocol from \cite{comp_self_testing}] \label{protocol:comp_self_testing}
~
\begin{enumerate}
\item Alice chooses a basis (called the \emph{state basis}) $\theta^A \in \{\texttt{Computational, Hadamard}\}$ uniformly at random and generates a key $k^A$ together with a trapdoor $t^A$, where the generation procedure for $k^A$ and $t^A$ depends on the the state basis $\theta^A$ and a security parameter $\lambda$.
Likewise, Bob generates $\theta^B, k^B$, and $t^B$.
The keys are such that the device cannot efficiently compute the state bases $\theta^A, \theta^B$ from the keys $k^A, k^B$.
Alice and Bob send the keys $k^A, k^B$ to the device.
\label{step1}
\item Alice and Bob receive strings $c^A$ and $c^B$, respectively, from the device. \label{step2}
\item[] \emph{Honest behaviour: Prepare a product state $\ket{\psi^A} \ket{\psi^B}$, where $\ket{\psi^A}$ is a function of the key $k^A$, and $\ket{\psi^B}$ is a function of $k^B$. Measure part of $\ket{\psi^A}$ to obtain a string $c^A$ and send it to Alice, keeping the remainder of the state. Similarly, obtain $c^B$ from $\ket{\psi^B}$ and send it to Bob.}
\item Using shared randomness, Alice and Bob choose a \emph{challenge type} $\ct \in \{\texttt{a}, \texttt{b}\}$ uniformly at random and send it to the device. \label{step3}
\end{enumerate}
If $\ct = \texttt{a}$:
\begin{enumerate}[label=\emph{a\arabic*}., ref=a\arabic*, resume]
\item Alice and Bob receive strings $z^A$ and $z^B$, respectively, from the device. \label{stepa4}
\item[] \emph{Honest behaviour: Measure the remainder of the states $\ket{\psi^A}$ and $\ket{\psi^B}$ in the computational basis and send back the resulting strings $z^A$ and $z^B$ to Alice and Bob, respectively.}
\end{enumerate}
If $\ct = \texttt{b}$:
\begin{enumerate}[label=\emph{b\arabic*}., ref=b\arabic*, resume]
\setcounter{enumi}{3}
\item Alice and Bob receive strings $d^A$ and $d^B$, respectively, from the device. \label{stepb4}
\item[] \emph{Honest behaviour: Measure the remainder of the states $\ket{\psi^A}$ and $\ket{\psi^B}$, except for one qubit of each state, in the Hadamard basis and send back the resulting strings $d^A$ and $d^B$ to Alice and Bob, respectively.}
\item Alice and Bob choose uniformly random \emph{measurement bases} (\emph{questions}) $x, y \in \{\texttt{Computational, Hadamard}\}$, respectively, and send them to the device.
\item Alice and Bob receive answer bits $a$ and $b$, respectively, from the device.\footnote{For readers familiar with QKD, we point out that these bits $a$ and $b$ (after many sequential repetitions) will form the basis for the raw key in our QKD protocol, \cref{protocol:comp_qkd}.} \label{step:self_testing_last}
\item[] \emph{Honest behaviour: The remaining state has two qubits. 
Apply a controlled-Z operation between them, followed by a Hadamard gate on the second qubit.
Measure the first qubit in basis $x$ and the second in basis $y$, obtaining outcomes $a, b \in \bits$, respectively. Send answer $a$ to Alice and $b$ to Bob.}
\end{enumerate}
\end{protocol}

In analogy to self-testing, we need to define what it means for the device to win the game.
This is done by specifying a number of checks that Alice and Bob apply to the device's answers.
These checks are described in \cref{app:honest_behaviour} and~\cite{comp_self_testing}. 
Here, we only remark that the computationally efficient evaluation of these checks requires the trapdoors~$t^A$ and~$t^B$, which are known to Alice and Bob, but not the device.

To state the self-testing guarantee from \cite{comp_self_testing}, we need a bit of notation.
The reference states (i.e., the states that the device is meant to prepare) in \cref{protocol:comp_self_testing} are Bell states.
We denote the four Bell states by 
\begin{equation}
\ket{\phi^{(s^A, s^B)}} = (\sigma_Z^{s^A} \sigma_X^{s^B} \ot \1) \frac{\ket{00} + \ket{11}}{\sqrt{2}} \label{eqn:bell_states_def}
\end{equation}
for bits $s^A, s^B$.
The reference measurements are single-qubit measurements in the computational or Hadamard basis.
We denote by $\{Q_x^a\}_{a \in \bits}$ the single-qubit measurement in the basis $x$ (e.g., for $x = \texttt{Hadamard}$, $Q^{0}_{x} = \proj{+}$ and $Q^{1}_{x} = \proj{-}$).

We are interested in the device's state and measurements in the last step of the protocol (\cref{step:self_testing_last}) in the case where $\theta^A = \theta^B = \texttt{Hadamard}$.
In this case, we denote the device's state by $\sigma^{(s^A, s^B)}$,
where $s^A$ and $s^B$ are bits that label which of the four Bell states the device should have prepared.
Alice can efficiently compute the bit $s^A$ from $k^A, t^A, c^A, d^A$, and likewise Bob can compute $s^B$ (see \cref{eqn:compute_s} for details).
In contrast, the device cannot \emph{efficiently} compute $s^A$ or $s^B$ because it does not have access to the trapdoors $t^A$ and $t^B$.
Hence, Alice and Bob know which Bell state the device should have prepared, whereas the device itself does not.

For questions $x, y \in \{\texttt{Computational, Hadamard}\}$, we denote the 4-outcome measurement used by the device to obtain answers $a, b \in \bits$ by $\{P^{(a, b)}_{x, y}\}_{a, b \in \bits}$.
Note that any arbitrary device that returns $a, b$ can always be described as performing a measurement on a state, so these definitions impose no additional assumptions on the device.
With this, we can state the self-testing guarantee from \cite{comp_self_testing} (in a simplified form).

\begin{theorem}[Theorem 4.38 in \cite{comp_self_testing}, simplified] \label{thm:self_testing}
Consider a device that wins \cref{protocol:comp_self_testing} with probability $1 - \eps$ and make the LWE assumption.
Let $\lambda$ be the security parameter used in the protocol, $s^A$ and $s^B$ bits (labelling the desired Bell state, as explained above), $\H$ the device's physical Hilbert space, and~$\H'$ some ancillary Hilbert space.
Then, there exists an isometry $V: \H \to \C^4 \ot \H'$ and some state $\xi^{(s^A, s^B)}_{\H'}$ such that, in the case $\theta^A = \theta^B = \textnormal{\texttt{Hadamard}}$, the following holds (with $\ket{\phi^{(s^A, s^B)}}$ and $Q_x^a$ as defined in and below \cref{eqn:bell_states_def}):
\begin{multline}
V P^{(a, b)}_{x, y} \sigma^{(s^A, s^B)} P^{(a, b)}_{x, y} V^\dagger  \approx_{O(\eps^c) + \negl(\lambda)} \\
\left( (Q^{a}_{x} \ot Q^{b}_{y}) \proj{\phi^{(s^A, s^B)}}  (Q^{a}_{x} \ot Q^{b}_{y}) \right) \ot \xi^{(s^A, s^B)}_{\H'} \,. \label{eqn:self_test_security}
\end{multline}
Here, the notation $\approx_{O(\eps^c) + \negl(\lambda)}$ means that the trace distance between the two states is $O(\eps^c) + \negl(\lambda)$ for some small constant~$c$ arising in the proof.
\end{theorem}

Intuitively, \cref{thm:self_testing} states that up to a change of basis (given by the isometry), any computationally bounded device that succeeds in the protocol must have performed single-qubit measurements on a Bell pair to obtain the results returned to the verifier.

We conclude this section with some intuition as to why \cref{thm:self_testing} holds. Depending on Alice's and Bob's choices for the state bases and challenge type, we distinguish two types of rounds: we call rounds with $\theta^A = \theta^B = \texttt{Hadamard}$ and challenge type \texttt{b} \emph{Bell rounds}, and all other rounds \emph{product rounds}.

\cref{thm:self_testing} only makes a statement about the device in a Bell round. 
In a Bell round, the two qubits prepared by an honest device at the start of \cref{step:self_testing_last} are Hadamard basis states (see \cref{app:honest_behaviour} for details). 
An honest device will apply a controlled-$Z$ gate followed by a single-qubit Hadamard gate to these two qubits, creating a Bell state $\sigma^{(s^A, s^B)} = \proj{\phi^{(s^A, s^B)}}$. 

The product rounds are used to check that the device behaves honestly.
In a product round, at least one of the two qubits prepared by an honest device at the start of \cref{step:self_testing_last} is in a computational basis state.
Thus, the two qubits will remain in a product state even after the controlled-$Z$ operation.
In a product round, the checks that the device needs to pass are independent for Alice's and Bob's components.
Intuitively, this implies that to pass the checks, any device needs to treat Alice's and Bob's components separately, i.e., always keep a product state in its register.
Further, one can show that the checks in a product round also ensure that the device has prepared the \emph{correct} product state.

Recall from \cref{step1} that a computationally bounded device does not know the bases $\theta^A$ and $\theta^B$, and therefore does not know whether it is in a Bell round or a product round.
To succeed in a product round with high probability, the device needs to behave honestly and prepare the correct single-qubit states.
Since the device cannot distinguish between the two round types, one can show that the device also needs to prepare the correct single-qubit states, i.e., Hadamard basis states, at the start of \cref{step:self_testing_last} in a Bell round.
An additional check in Bell rounds ensures that the controlled-$Z$ operation has  been applied correctly on these states, creating a Bell pair.
The full security proof can be found in~\cite{comp_self_testing}.\footnote{The proof in~\cite{comp_self_testing} proceeds differently from the intuition given above; the interested reader is directed to~\cite{comp_self_testing}.}

\section{Key distribution protocol \label{sec:comp_qkd}}

\subsection{Main ideas}

We are now ready to describe our DIQKD protocol, \cref{protocol:comp_qkd} below.
The main building block of our DIQKD protocol is the self-testing protocol, \cref{protocol:comp_self_testing}, introduced in the previous section. 
On a high level, the idea is the following: Alice and Bob each receive a component of the key generation device and execute $n$ rounds (in sequence) of \cref{protocol:comp_self_testing} (with some modifications, see \cref{protocol:self_testing_teleportation} below), collecting the devices' inputs and outputs for each round.
Then, they use the observed input-output behaviour in a subset of the $n$ rounds to calculate the proportion of rounds that satisfy the winning condition of \cref{protocol:comp_self_testing}. 
If they find that the device wins a sufficiently high proportion of these test rounds, they can use the device's output in the remaining rounds to generate a secure key.
The security of this key is based on \cref{thm:self_testing}, which certifies the states and measurements used by the device.
This certification replaces the usual Bell-based certification of the device.

Note that in contrast to the self-testing setting, in the DIQKD setting the state of the computationally bounded device may additionally be entangled with the computationally unbounded adversary Eve.
However, because Eve can only act on her part of the state and not assist the device in breaking the computational assumption, we can still apply \cref{thm:self_testing} to the reduced state of the device.
Then, \cref{thm:self_testing} asserts that, after applying an isometry, the device's state is a Bell state tensored with some additional state, and that the device's measurements only act on the Bell state.
Hence, the additional state is irrelevant for the measurement outcomes.
Because the device's Bell state is a pure state, it cannot be entangled with the state of the adversary, so the adversary's marginal does not reveal any information about the device's measurement outcomes.
We will additionally need to ensure that the classical information exchanged by Alice and Bob during the parameter estimation phase of \cref{protocol:comp_qkd} does not leak any information to the adversary, which is covered in detail in the proof of \cref{thm:security_iid}.

As mentioned in \cref{sec:intro}, we would like the \emph{honest} device to be able to succeed using only EPR pairs and local operations.
This requires a modification to \cref{protocol:comp_self_testing} because the honest behaviour in \cref{step:self_testing_last} of \cref{protocol:comp_self_testing} uses a non-local controlled-$Z$-gate between Alice's and Bob's components of the device.
We can remove the need for this non-local operation using \emph{gate teleportation} with pre-shared EPR pairs~\cite{gate_teleportation}.
For example, consider the following circuit (adapted from~\cite{childs2005unified}) that only uses one EPR pair and local operations:
\begin{center}
\vspace{-20pt}
\begin{equation} \label{eqn:teleport_circ}
\begin{quantikz}[row sep={0.1cm}]
\lstick{$A$} & \qw & \targ{} & \meter{} \rstick{$h^A$} \\
\lstick[wires=2]{$\ket{\text{EPR}}$} & \qw & \ctrl{-1} & \qw \\
& \gate{H} & \ctrl{1} & \qw \\
\lstick{$B$} & \qw & \targ{} & \meter{} \rstick{$h^B$}
\end{quantikz}
\end{equation}
\end{center}

If the initial state on registers $A$ and $B$ is $\ket{\psi}_{AB}$, the output state on the middle two registers is 
\begin{equation}
\sigma_X^{h^A} \sigma_Z^{h^B} \ot \sigma_X^{h^B} \sigma_Z^{h^A} CZ \ket{\psi}_{AB} \,, \label{eqn:teleportation_outcome}
\end{equation}
where $CZ$ is the controlled-$Z$ gate.
Hence, we have applied the desired controlled-$Z$ operation to $\ket{\psi}_{AB}$, followed by an additional ``correction operator'' $\sigma_X^{h^A} \sigma_Z^{h^B} \ot \sigma_X^{h^B} \sigma_Z^{h^A}$.

An honest device for \cref{protocol:comp_self_testing} still needs to undo the correction operator $\sigma_X^{h^A} \sigma_Z^{h^B} \ot \sigma_X^{h^B} \sigma_Z^{h^A}$.
In principle, the two components of the device could communicate the bits $h^A$ and $h^B$ to each other and apply local operations that cancel this correction operator.
However, as explained above, we want an honest device to be able to succeed without communication between its components.

Instead of the honest device having to deal with the correction operators, we can modify the checks of \cref{protocol:comp_self_testing}.
For this, note that the honest strategy in \cref{protocol:comp_self_testing} measures the state in the computational or Hadamard basis immediately after applying the controlled-$Z$ operation (up to applying a single-qubit Hadamard gate, which only relabels measurement bases and which we ignore here for simplicity).
The correction operator only switches these measurement outcomes.
Hence, the honest device can return the bits $h^A$ to Alice and $h^B$ to Bob in addition to the measurement outcomes $a$ and $b$.
Alice and Bob then use their authenticated classical communication to undo the effect that the correction operator $\sigma_X^{h^A} \sigma_Z^{h^B} \ot \sigma_X^{h^B} \sigma_Z^{h^A}$ had on the device's measurement outcomes.
Furthermore, it turns out that if we are in the Bell case in \cref{protocol:comp_self_testing}, the state prepared by the honest device is an eigenstate of the correction operator, so in this case, Alice and Bob do not need to perform any correction on the device's reported outcomes (see \cref{app:honest_behaviour} for details).

\subsection{Formal protocol description and security analysis}

We now describe our QKD protocol in more detail. First, we give a modified version of \cref{protocol:comp_self_testing} adapted to the use of gate teleportation by an honest device as described above.\footnote{In addition, we also change how Alice and Bob sample the challenge type. In \cref{protocol:comp_self_testing}, they use shared randomness for this purpose. This would still work in the QKD setting (where they would use public classical communication to establish the shared randomness), but it slightly simplifies the security analysis to assume that they sample challenge types independently and then post-select on having sampled the same challenge type. The downside of this is a reduction of the key rate in \cref{thm:security_iid} by a constant factor of 2.}

\begin{protocol}[Modified self-testing protocol] \label{protocol:self_testing_teleportation}
\changed{Alice and Bob execute the same steps as in \cref{protocol:comp_self_testing}, with the following modifications.
In \cref{step3}, challenge types $\ct^A$ and $\ct^B$ are sampled independently by Alice and Bob, respectively, and sent to the device; and in \cref{step:self_testing_last}, Alice additionally receives a bit $h^A$ and Bob a bit $h^B$.
The remaining steps are as in \cref{protocol:comp_self_testing}, where now Alice acts according to $\ct^A$ and Bob according to $\ct^B$: for example, if $\ct^A = \texttt{a}$ and $\ct^B = \texttt{b}$, then Alice will receive a string $z_A$ as in \cref{stepa4}, and Bob will execute Steps \ref{stepb4}-\ref{step:self_testing_last} (with the above modification, i.e.~he will receive a bit $h^B$ in \cref{step:self_testing_last}).

\emph{Honest behaviour: In the case $\ct^A = \ct^B$, behave as in \cref{protocol:comp_self_testing}, but in \cref{step:self_testing_last}, use a pre-shared EPR pair and gate teleportation to apply the controlled-$Z$ operation as in \cref{eqn:teleport_circ} and additionally return the bits $h^A, h^B$ from the gate teleportation to Alice and Bob, respectively.
Note that in this strategy, the actions of Alice's and Bob's side of the device are independent.
Hence, we can extend the honest strategy to the case $\ct^A \neq \ct^B$ with each side individually acting according to the challenge type it has received.}}
\end{protocol}

Like \cref{protocol:comp_self_testing}, this protocol depends (implicitly) on a security parameter $\lambda$.
Both the honest behaviour and the winning condition used by Alice and Bob are described in more detail in \cref{app:honest_behaviour}.
Note that if we adapt the checks from \cref{protocol:comp_self_testing} accordingly, \cref{thm:self_testing} still applies to \cref{protocol:self_testing_teleportation} since any device that could cheat in \cref{protocol:self_testing_teleportation} (where also $h^A$ and $h^B$ are returned) can easily be converted into a device that cheats in the original protocol.

Our key distribution protocol below executes $n$ rounds of \cref{protocol:self_testing_teleportation}, then uses classical communication to estimate the proportion of rounds satisfying the winning condition, and extracts a secure key using standard classical post-processing steps, namely classical error correction and privacy amplification.
Recall that the setting for this protocol is that of \cref{fig:setting_comp}, i.e., Alice and Bob each receive a component of a device prepared by the adversary, and the two components can be connected by a quantum channel.

\begin{protocol}[Key distribution protocol] \label{protocol:comp_qkd}
~\\
\textbf{Parameters:}
\begin{itemize}[label={--}, noitemsep, topsep=0pt]
\item number of rounds $n$,
\item threshold for tolerated losing frequency $\eps$,
\item security parameter $\lambda$.
\end{itemize}
\textbf{Data generation:}
\begin{enumerate}
\item \label{step:generate_data} Alice and Bob execute $n$ rounds of \cref{protocol:self_testing_teleportation} (with security parameter $\lambda$).
For each round $i \in \{1, \dots, n\}$, Alice stores the following classical information: 
\begin{itemize}[label={--}, noitemsep, topsep=-2pt]
\item the state basis $\theta_i^A$,
\item the key $k_i^A$,
\item the trapdoor $t_i^A$,
\item the string $c_i^A$,
\item the challenge type $\ct_i^A \in \{\texttt{a}, \texttt{b}\}$, and
\item either the string $z_i^A$ if $\ct_i^A =$\texttt{a},
\item or the tuple $(d_i^A, x_i, a_i, h_i^A)$ if $\ct_i^A=$\texttt{b}.
\end{itemize}
Analogously, Bob stores $\theta_i^B, k_i^B, t_i^B, c_i^B$; and $z_i^B$ if $\ct_i^B =$\texttt{a}, or $(d_i^B, y_i, b_i, h_i^B)$ if $\ct_i^B =$\texttt{b}.
\item For every $i \in \{1, \dots, n\}$, Alice and Bob each publish their state bases $\theta_i^A, \theta_i^B$ and challenge types $\ct_i^A, \ct_i^B$ via their authenticated public channel.
They both store a variable $\rt_i$ (indicating the round type), defined as follows:
\begin{itemize}[label={--}, noitemsep, topsep=-2pt]
\item if $\ct_i^A \neq \ct_i^B$, set $\rt_i = \perp$;
\item else if $\ct_i^A = \ct_i^B = \texttt{b}$ and $\theta_i^A = \theta_i^B = \texttt{Hadamard}$, set $\rt_i = \texttt{Bell}$;
\item else, set $\rt_i = \texttt{Product}$.
\end{itemize}
\item For every $i \in \{1, \dots, n\}$, Alice chooses $T_i$, indicating a test round or generation round, as follows:
\begin{itemize}[label={--}, noitemsep, topsep=-2pt]
\item if $\rt_i = \texttt{Bell}$, choose $T_i \in \{\texttt{Test}, \texttt{Generate}\}$ uniformly at random;
\item else, set $T_i = \texttt{Test}$.
\end{itemize}
Alice publishes $(T_1, \dots, T_n)$ (so Bob also has access to it).
\end{enumerate}
\textbf{Sifting:}
\begin{enumerate}[resume, leftmargin=11pt, topsep=0pt]
\item Alice and Bob discard all rounds with $\rt_i = \perp$. Let $n'$ be the number of remaining rounds (re-indexed as $1, \dots, n'$). \label{step:sift}
\end{enumerate}
\textbf{Parameter estimation:}
\begin{enumerate}[resume]
\item For every $j \in \{1, \dots, n'\}$ with $T_j = \texttt{Test}$, Bob publishes his entire inputs and outputs from round $j$ (described in \cref{step:generate_data}). 
Using Bob's published data, Alice uses \cref{eqn:psi_3} to compute which bits $a_j', b_j'$ the honest device would have returned (for the given values of $h^A, h^B$).
If $a_j \neq a_j'$ or $b_j \neq b_j'$, she sets a variable $W_j$ to $\texttt{fail}$.
\item Alice computes the fraction of (sifted) test rounds where $W_j = \texttt{fail}$. If this exceeds $\eps$, the protocol aborts.
\end{enumerate}
\textbf{Key extraction:}
\begin{enumerate}[resume]
\item For every $j \in \{1, \dots, n'\}$ with $T_j = \texttt{Generate}$ (which, by definition of $T_j$, is also a Bell round), Alice and Bob compute the bits $s_j^A$ and $s_j^B$, respectively, using \cref{eqn:compute_s}. 
They publish their measurement bases $x_j, y_j$. 
If $x_j = y_j = \texttt{Computational}$, then Bob sets $\tilde{b}_j = b_j \oplus s_B$ (while Alice keeps her bit $a_j$ unchanged).
Otherwise, they set $a_j = b_j = \perp$ (no key can be generated). \label{step7qkd}
\item Alice and Bob apply one-way error correction and privacy amplification to their strings $A = a_1 \dots a_{n'}$ and $B = \tilde{b}_1 \dots \tilde{b}_{n'}$ to generate their key.
\end{enumerate}
\end{protocol}

An honest device will simply execute the honest behaviour for \cref{protocol:self_testing_teleportation} identically and independently in each round (see \cref{app:honest_behaviour} for details).
As explained above, this only requires pre-shared EPR pairs and local operations.

Our goal is to prove \cref{thm:security_iid}: assuming that the device does not break the LWE assumption, we need to show that our protocol's key rate $K$ is at least $\frac{1}{128} - O(\eps^c \log \eps) - \negl(\lambda)$, where $c$ is the same constant as in \cref{thm:self_testing}.

The outline of the proof is as follows:
we first define the state $\rho$ that contains Alice and Bob's classical information at the end of the protocol, as well as Eve's quantum side information.
This state is the result of measurements that the device performed on its state.
Because the self-testing protocol (\cref{protocol:comp_self_testing}) gives us control over both the device's state and measurements, we can apply \cref{thm:self_testing} to show that the state $\rho$ is close to some \emph{ideal state} $\tilde{\rho}$, and that the device measured $\tilde{\rho}$ in the requested bases.
This ideal state $\tilde{\rho}$ is essentially the final state of executing \cref{protocol:comp_qkd} with an honest device.
We then show that the ideal state $\tilde{\rho}$ leads to a key rate of at least $\frac{1}{128}$. 
For this, we need to show that the classical information publicly communicated between Alice and Bob in \cref{protocol:comp_qkd} does not reveal too much information about the secret key to Eve.
Finally, writing the bound on the key rate of the ideal state $\tilde{\rho}$ in terms of conditional entropies, we can derive a lower bound on the key rate of the actual state $\rho$ using the closeness of $\tilde{\rho}$ and $\rho$ and a continuity bound on the conditional entropy from \cite{winter2016tight}.

\begin{proof}[Proof of \cref{thm:security_iid}] \label{security_prf}
The device used by Alice and Bob is prepared by the adversary Eve.
Hence, if the initial state of the device is $\psi_{AB}$, Eve can hold a purifying system, so that the system as a whole is described by $\ket{\psi_{ABE}}$.\footnote{Assuming that an arbitrary Eve holds the purifying system is without loss of generality, as the purifying system gives the maximal amount of information about the state $\psi_{AB}$.}

Consider the state at the end of \cref{step7qkd}: because we are making the IID assumption, this state is an $n'$-fold tensor product of a state $\rho_{XYABTOE}$.
Each of the $n'$ copies of $\rho$ corresponds to one of the $n'$ rounds of the protocol (after the sifting step) and contains the following registers:
\begin{itemize}[label={--}, noitemsep, topsep=0pt]
\item $X, Y$ and $A, B$ are classical random variables for Alice's and Bob's questions and answers, respectively.
\item $T$ is a classical random variable indicating a test round ($T = \texttt{Test}$) or a generation round ($T = \texttt{Generate}$).
\item $O$ is a classical random variable containing the remaining information that Alice and Bob publish in a round of the protocol.
For test rounds, this comprises the entire interaction between Alice, Bob, and the device (i.e., the information listed in \cref{step:generate_data}).
For generation rounds, the state bases, challenge types, and measurement bases are published. Conditioning on $T = \texttt{Generate}$ already fixes the state bases and challenge types. The questions $x, y$ are stored in registers $X$ and $Y$.
Therefore, the register $O$ is empty if $T = \texttt{Generate}$.
\item $E$ contains Eve's quantum side-information.
\end{itemize}
Note that Alice and Bob also hold additional private information in generation rounds (such as the bits $s^A$ and $s^B$), but this information can be discarded and is not included in $\rho_{XYABTOE}$.

\cref{protocol:comp_qkd} applies one-way error correction and privacy amplification to the raw key in registers $A$ and $B$ of $\rho_{XYABTOE}^{\ot n'}$.
Therefore, the key rate $K_{\rho}$ achieved by our protocol in the limit $n \to \infty$ and under the IID assumption is lower-bounded by~\cite{dw, renner2005simple}:
\begin{equation*}
K_{\rho} \geq \frac{1}{2} \left(H(A | E X Y O T)_{\rho} - H(A | B X Y O T)_{\rho}\right) \,,
\end{equation*}
where $H$ is the conditional von Neumann entropy.
The additional factor of $\frac{1}{2}$ arises because half the rounds are sifted out in \cref{step:sift}.

We can split this expression according to the round type:
\begin{multline*}
K_{\rho} \geq \frac{1}{2} \sum_{t \in \{\texttt{Test,Gen.}\}} \pr{T = t} \Big( H(A | E X Y O, T=t)_{\rho} \\ - H(A | B X Y O, T=t)_{\rho} \Big) \,,
\end{multline*}
where $\pr{T = \texttt{Test}} = 1 - \frac{1}{16}$ and $\pr{T = \texttt{Generate}} = \frac{1}{16}$ are the probabilities of choosing a test and a generation round, respectively (conditioned on the round not having been sifted out).

In a test round, Alice and Bob publish their entire inputs and outputs (now stored in register $O$), including $a$ and $b$. Hence, the conditional entropies for $T = \texttt{Test}$ are both $0$.

We now turn to the analysis of a generation round.
We denote by $s^A$ and $s^B$ the bits computed by Alice and Bob in \cref{step7qkd}. 
For a single round of the data generation step (\cref{step:generate_data} of \cref{protocol:comp_qkd}), let $\sigma^{(s^A, s^B)}_{A'B'E}$ be the joint state of the device and Eve's side information right before the device performs the measurements $\left( P^{(a, b)}_{x, y} \right)_{A'B'}$ (corresponding to the state before \cref{step:self_testing_last} of \cref{protocol:comp_self_testing}, with the notation introduced for \cref{thm:self_testing}).
Here, $A'$ is the (quantum) register of Alice's component of the device, $B'$ is the (quantum) register of Bob's component, and $E$ contains Eve's quantum side information.

Using the same notation as for \cref{protocol:comp_self_testing}, the state of Alice's and Bob's question and answer registers as well as Eve's quantum side information after a single round of \cref{step:generate_data} in \cref{protocol:comp_qkd} is
\begin{multline}
\sum_{x, y, a, b} \frac{1}{4} {\rm Tr}_{A'B'} \left[ \left( P^{(a, b)}_{x, y} \right)_{A'B'} \sigma^{(s^A, s^B)}_{A'B'E} \left( P^{(a, b)}_{x, y} \right)_{A'B'}  \right] \\
\ot \proj{x, y, a, b}_{XYAB} \,.
\end{multline}
The factor of $\frac{1}{4}$ arises because Alice and Bob choose the questions $x, y \in \{\texttt{Computational}, \texttt{Hadamard}\}$ uniformly at random.
Note that because we allow arbitrary quantum communication between Alice's and Bob's components of the device, the device's measurements $\left( P^{(a, b)}_{x, y} \right)_{A'B'}$ could be global measurements, not just product measurements $( P^{a}_{x} )_{A'} \ot ( P^{b}_{y} )_{B'}$.

Now observe that the checks applied by Alice and Bob in a test round are equivalent to the checks applied in the (modified) self-testing protocol, \cref{protocol:self_testing_teleportation}.
Since we are considering the asymptotic IID case with $n \to \infty$, if the protocol does not abort for threshold $\eps$, this means that the winning condition from \cref{protocol:self_testing_teleportation} must be satisfied with probability at least $1 - \eps$ in test rounds.
At the end of \cref{step:generate_data}, it has not yet been decided whether a particular round will be a test or a generation round.
Hence, \cref{thm:self_testing} also applies to the state and measurements in a generation round.

Applying \cref{thm:self_testing} to the state $\sigma^{(s^A, s^B)}_{A'B'E}$, and using the continuity and cyclicity of the trace as well as $V^\dagger V = \1$, we find that the physical state $\rho$ at the end of \cref{step7qkd} must be within trace distance $O(\eps^c) + \negl(\lambda)$ of the \emph{ideal state}
\begin{multline}
\tilde{\rho} = \sum_{s^A, s^B, x, y, a, b \in \bits} \frac{1}{4} \; p_{x, y}^{a, b, s^A, s^B} \; \rho^{(s^A, s^B)}_{E} \\
\ot \proj{x, y, a, \tilde{b}}_{XYAB} \,, \label{eqn:state_whole}
\end{multline}
where
\begin{equation*}
p_{x, y}^{a, b, s^A, s^B} = \bra{\phi^{(s^A, s^B)}} Q^{a}_{x} \ot Q^{b}_{y} \ket{\phi^{(s^A, s^B)}} \,,
\end{equation*}
and
\begin{equation*}
\rho^{(s^A, s^B)}_{E} = {\rm Tr}_{A'B'} \left[ \xi^{(s^A, s^B)}_{A'B'E} \right]
\end{equation*}
is Eve's quantum side information. 
Here, we have used the same notation as in \cref{thm:self_testing}, and $\tilde{b} = b \oplus s_B$ as in \cref{step7qkd}.
As explained above, the side information register $O$ is empty in a generation round.

We now analyse the key rate of the ideal state $\tilde{\rho}$.
If $x = y = \texttt{Computational}$ does not hold, the key rate is $0$ because Alice and Bob both set their output registers to $\perp$.

Conditioned on $x = y = \texttt{Computational}$, the measurement outcomes for the ideal state $\tilde{\rho}$ are either perfectly correlated or perfectly anti-correlated, depending on $s^B$. 
Since Bob flips his bit $b$ to get $\tilde{b}$ in the anti-correlated case, we always have $a = \tilde{b}$.
In other words, $p_{x, y}^{a, b, s^A, s^B} = 1$ if $a = \tilde{b}$, and $p_{x, y}^{a, b, s^A, s^B} = 0$ otherwise.
Therefore, in this case we have (with $x = \texttt{Computational}$)
\begin{multline*}
\tilde{\rho} = \sum_{\substack{s^A, s^B, a \in \bits}} \frac{1}{4} \; \rho^{(s^A, s^B)}_{E}
\ot \proj{x, x, a, a}_{XYAB}  \,.
\end{multline*}

Note that in this expression, the sum over $a$ is independent of the rest. 
Hence, the state is in a product between the registers $AB$ and the remaining registers. 
Therefore, in the calculation of the key rate, conditioning on the remaining registers does not change the entropy and we have that for the \emph{ideal state} $\tilde{\rho}$: 
\begin{multline}
K_{\tilde{\rho}} = \frac{\pr{T = \texttt{Gen.}}}{8} \big( H(A|X = Y = \texttt{Comp.}, T = \texttt{Gen.})_{\tilde{\rho}} \\
- H(A|B, X = Y  = \texttt{Comp.}, T = \texttt{Gen.})_{\tilde{\rho}}  \big) \,. \label{eqn_dw_ideal}
\end{multline}
The additional factor of $\frac{1}{4}$ arises from the conditioning on $X = Y = \texttt{Computational}$.
Conditioned on $X = Y= \texttt{Computational}$ and $T = \texttt{Generate}$, the value of $A$ is uniformly random, and the value of $B$ equals that of $A$.
Hence, 
\begin{equation*}
H(A|X = Y = \texttt{Comp.}, T = \texttt{Gen.})_{\tilde{\rho}} = 1
\end{equation*}
and 
\begin{equation*}
H(A|B, X = Y = \texttt{Comp.}, T = \texttt{Gen.})_{\tilde{\rho}} = 0 \,.
\end{equation*}
Plugging this into \cref{eqn_dw_ideal}, we obtain
\begin{equation*}
K_{\tilde{\rho}} = \frac{\pr{T = \texttt{Gen.}}}{8} = \frac{1}{128} \,.
\end{equation*}

This is the key rate for the ideal state $\tilde{\rho}$.
However, we are interested in the key rate for the state $\rho$ that the device actually uses in the protocol.
To connect the two, recall that by \cref{thm:self_testing}, the two states differ by at most $O(\eps^c) + \negl(\lambda)$ in trace distance.
Therefore, we can apply a continuity bound for the conditional entropy from \cite{winter2016tight}.
Using the fact that the classical registers have a fixed constant dimension and absorbing the resulting constant from \cite{winter2016tight} into the $O$-notation, we find that 
\begin{equation*}
K_{\rho} = \frac{1}{128} - O(\eps^c \log \eps) - \negl(\lambda) \,.
\end{equation*}
This completes the proof of \cref{thm:security_iid}.
\end{proof} 

As we noted when we stated \cref{thm:security_iid}, the constant $\frac{1}{128}$ is a consequence of fixing certain parameters in the protocol to $\frac{1}{2}$.
Specifically, from the proof, we see that this constant arises from the probability $\frac{1}{32}$ of being in a generation round, and the probability $\frac{1}{4}$ of choosing $x = y = \texttt{Computational}$.
For practical applications, these probabilities could be optimized and treated as functions of the number of rounds $n$ to increase the key rate to up to $1 - O(\eps^c \log \eps) - \negl(\lambda)$ in the asymptotic IID scenario.

\section{Discussion \label{sec:discussion}}

We have considered the question whether there are alternatives to the no-communication assumption used in standard Bell inequality-based DIQKD protocols.
For this, we have introduced a modified setting for DIQKD (see \cref{fig:setting_comp}) that drops the no-communication assumption and allows the two components of the key generation device to exchange quantum communication.
Instead, we have assumed that the key generation device is computationally bounded and cannot break the LWE assumption, a standard assumption in post-quantum cryptography.
For this setting, we have described a protocol that allows Alice and Bob to generate an \emph{information-theoretically} secure key and shown that it achieves a  positive key rate.

Unlike previous approaches to weakening the no-communication assumption~\cite{silman2013device, tavakoli2019informationally, tavakoli2020characterising}, which required an a priori device-\emph{dependent} upper bound on the amount of information exchanged between different parts of the device, the LWE assumption is a general assumption about any computationally bounded quantum device, and our belief in it does not require us to inspect the specific device at hand in detail.

As noted in \cref{sec:intro}, our modified DIQKD setting allows for arbitrary quantum communication between the components of the device, but requires that the adversary cannot access this communication channel.
While a private channel between the device components is a strictly weaker assumption than the no-communication assumption, in practice, the privacy of the channel connecting the two components may be as hard to ensure as the original no-communication assumption, or the assumption that Eve can send EPR states to the device via a strictly one-way channel.
The setting in \cref{fig:setting_comp} should therefore be viewed as an extreme case meant for studying DIQKD without the no-communication assumption.
An actual implementation of DIQKD could adopt a multi-layered approach: physical shielding of the device gives us some credence in the no-communication assumption, but we might still want to employ the protocol we developed in this paper to further boost our confidence in the security of the final key.

With this approach in mind, it is crucial that the behaviour of an honest device for any protocol developed for the setting in \cref{fig:setting_comp} can also be executed in the standard setting, i.e., that an honest device can succeed with local operations and pre-shared EPR pairs.
Our \cref{protocol:comp_qkd} satisfies this requirement.
If one drops this requirement and further assumes that Alice has access to trusted \emph{private} randomness (whereas standard DIQKD and our protocol only require public randomness), then one could execute the following simple key distribution protocol: Alice inputs a random string into her component of the device and asks the device to output this string at Bob's end. Alice and Bob then publicly compare their strings at a subset of locations. If their strings agree, they can use the remainder of the shared string as a key.\footnote{We thank Carl Miller for pointing out this protocol to us.} While this protocol can, strictly speaking, be executed in the setting of \cref{fig:setting_comp}, there is no way for an honest device to succeed \emph{without} access to the communication channel connecting its components. Accordingly, this protocol cannot be used as part of the multi-layered approach to closing possible loopholes described in the previous paragraph.

One conceptually interesting aspect of the setting in \cref{fig:setting_comp} is that while it relies on a computational assumption, the resulting key is information-theoretically secure, just as in the standard DIQKD setting.
This means that even if the computational assumption is broken in the future, encrypted messages remain private, in contrast to classical public-key cryptography.
This ``lifting'' of a computational assumption to an information-theoretic guarantee appears to be a uniquely quantum capability~\cite{randomness}.

As noted in \cite{randomness}, the root of this ``quantum advantage'' lies in the interactive nature of the protocol and the incompatibility of different quantum measurements.
On a high level, the combination of interactivity and incompatibility allows the device to correctly answer any \emph{one} of two questions (corresponding to the two challenge types in \cref{protocol:comp_qkd}), but never both simultaneously.
In contrast, a classical device that is able to answer any one of two questions is also able to answer both at the same time.\footnote{In cryptography, a common way to state this is that a classical device can be ``rewound'': this means that it can first be used to answer the first question, then reset to its state before answering the first question, and subsequently be used to also answer the second question. In the quantum case, if answering the first question requires a measurement, then the fact that a measurement is destructive prevents us from resetting the device to its previous state~\cite{graafrewinding}.}

In particular, this kind of quantum advantage differs from both Bell non-locality because it does not require a device with spatially separated components, and from quantum computational supremacy because it is independent of whether or not quantum computation is classically simulable.
Therefore, protocols with cryptographic assumptions such as ours may also yield new insights into what separates the capabilities of quantum and classical devices, and might lead to conceptually new quantum cryptographic capabilities \cite{poqk1, poqk2, vidick2020classical}.

There are several important directions for future work; we list a few.
Firstly, here we have only shown the security of our protocol in the asymptotic IID scenario.
The analysis should, of course, be extended beyond the IID setting.
A related protocol for randomness expansion has been analysed in the non-IID setting~\cite{randomness}, and we expect that an analogous analysis will work for our protocol, too.
The analysis in~\cite{randomness}, however, is highly technical and we hope that new techniques, similar to those used in DIQKD~\cite{entropy_accumulation}, can be developed to simplify the analysis of our protocol in the non-IID setting.

Another important task is to improve the dependence on $\eps$ in the key rate for our protocol to become practical. 
In particular, this means increasing the constant $c$ (which we estimate is currently smaller than $10^{-3}$).
One can either approach this by streamlining the analysis of the self-testing protocol \cite{comp_self_testing}, or by taking a more direct approach that shows a lower bound on the key rate without explicitly using a self-testing statement (which is stronger than necessary for DIQKD).
Additionally, to improve practicality, one should try to optimize the post-quantum cryptographic tools used in \cite{randomness, verification, rsp, comp_self_testing} for smaller quantum devices \cite{brakerski2020simpler}.

\begin{acknowledgments}
This work was done in part while all authors were visiting the Simons Institute for the Theory of Computing and while RAF was associated with the EECS department of the University of California, Berkeley.
We thank Charles Ci Wen Lim, Christopher Portmann, and Thomas Vidick for their helpful comments.
TM acknowledges support from the ETH Foundation through the Excellence Scholarship \& Opportunity Programme, and from the AFOSR project No. FA550-19-1-0202.
YD was supported by the Dutch Research Council (NWO/OCW), as part of the Quantum Software Consortium programme (project number 024.003.037).
AC was a Quantum Postdoctoral Fellow at the Simons Institute for the Theory of Computing supported by NSF QLCI Grant No. 2016245. AC was also supported by DARPA under agreement No.~HR00112020023. Any opinions, findings and conclusions or recommendations expressed in this material are those of the author(s) and do not necessarily reflect the views of the United States Government or DARPA.
RAF was supported by a research grant from the Center for New Scientists at the Weizmann Institute of Science, the Swiss National Science Foundation via the Postdoc.Mobility grant, the MURI Grant FA9550-18-1-0161, and ONR award N00014-17-1-3025.
\end{acknowledgments}

\appendix

\section{Detailed description of winning condition and honest behaviour} \label{app:honest_behaviour}
In this appendix, we give a more detailed description of the winning condition and behaviour of the honest device for the modified self-testing protocol, \cref{protocol:self_testing_teleportation}.
We describe the winning condition and honest behaviour together to make it clear why the winning condition is chosen as it is, and why an honest device wins with probability negligibly close to 1.
The honest behaviour of a device in our key distribution protocol, \cref{protocol:comp_qkd}, is simply to execute the honest behaviour for \cref{protocol:self_testing_teleportation} independently in every round of the protocol.

To describe the honest behaviour for \cref{protocol:self_testing_teleportation}, we need to introduce \emph{extended trapdoor claw-free function families} (ETCF families), a cryptographic primitive introduced in \cite{randomness, verification} that underlies the self-testing protocol from \cite{comp_self_testing}.
An ETCF family consists of two families of functions pairs, $\mf$ and $\mg$.
Function pairs $(f_{k, 0}, f_{k, 1}) \in \mf$ are indexed by keys $k \in \kf$ and are called \emph{claw-free}; function pairs $(f_{k, 0}, f_{k, 1}) \in \mg$ are indexed by keys $k \in \kg$ and are called \emph{injective}.
For the purposes of the honest implementation, the two most important properties are the following:
\begin{enumerate}[label=(\roman*), leftmargin=20pt]
\item For every key $k \in \kf$ (i.e., a key for a claw-free pair), the functions $f_{k, 0}$ and $f_{k, 1}$ are injective and have the same domain and image, so together, the pair $(f_{k,0}, f_{k, 1})$ can be viewed as a 2-to-1 function.\footnote{Two elements $x_0, x_1$ from the domain that satisfy $f_{k,0}(x_0) = f_{k,1}(x_1)$ are called a claw. The function pair is called  ``claw-free'' because such claws are computationally difficult to find, not because they do not exist.}
Given the key $k$, a quantum computer can efficiently evaluate the functions $f_{k, 0}$ and $f_{k, 1}$ in superposition (so intuitively, one can think of $k$ as a set of instruction for how to evaluate $f_{k,0}$ and $f_{k, 1}$).
\item For every key $k \in \kg$ (i.e., a key for an injective pair), the functions $f_{k, 0}$ and $f_{k, 1}$ are injective, have the same domain, but have disjoint images.
As above, given the key $k$, a quantum computer can efficiently evaluate the functions $f_{k,0}$ and $f_{k, 1}$ in superposition.
\end{enumerate} 
We denote the common domain of all function pairs by $\mx$, and the codomain by $\my$, and assume that both are sets of all bit strings of a fixed length.

For the security proof of \cref{protocol:self_testing_teleportation}, additional cryptographic properties of these functions are used. We do not describe them here and refer to \cite{randomness, verification, comp_self_testing} for details, but note that ETCF families with these cryptographic properties can be constructed from the standard post-quantum cryptographic assumption that the LWE problem~\cite{lwe} is hard to solve on a quantum computer.\footnote{In fact, one cannot quite construct ETCF families as we have described them here from the LWE problem, but only an approximate version (called extended \emph{noisy} trapdoor claw-free families). The only consequence of this for our work is that an honest device cannot satisfy the winning conditions for \cref{protocol:self_testing_teleportation} with probability 1, but only with probability $1 - \negl(\lambda)$, where $\lambda$ denotes the security parameter (intuitively, the length of the keys $k \in \kf \cup \kg$) and $\negl(\lambda)$ is a negligible function in $\lambda$, i.e., a function that decays faster than any inverse polynomial.
We will ignore this subtlety for the rest of this appendix.} 

We can now describe the honest behaviour for \cref{protocol:self_testing_teleportation}, following the steps of \cref{protocol:comp_self_testing}.
In \cref{step1}, Alice's component of the device is given a key $k^A \in \kf \cup \kg$.
Whether $k^A \in \kf$ or $k^A \in \kg$ is determined by the value of Alice's state basis $\theta^A$: $\theta^A = \texttt{Hadamard}$ corresponds to $k^A \in \kf$, and $\theta^A = \texttt{Computational}$ corresponds to $k^A \in \kg$.
The device uses this key to evaluate the function pair $(f_{k,0}, f_{k,1})$ on the uniform superposition over the domain (dropping a normalization factor): 
\begin{equation}
\ket{\psi^A_0} = \sum_{b \in \bits} \sum_{x \in \mx} \ket{b}\ket{x}\ket{f_{k, b}(x)} \,.
\end{equation}
The device then measures the last register to obtain the string $c^A$, which is returned to Alice.
Bob's component of the device does the same with the key $k^B$.

At this point, it is instructive to consider the post-measurement state. 
If $k^A \in \kf$, then there are unique $x_0^A, x_1^A \in \mx$ such that $f_{k^A, 0}(x_0^A) = f_{k^A, 1}(x_1^A) = c^A$, so the post-measurement state is (up to normalization): 
\begin{equation}
\ket{\psi^A_1} = \ket{0}\ket{x_0^A} + \ket{1}\ket{x_1^A} \,. \label{eqn:post_f}
\end{equation}
On the other hand, if $k^A \in \kg$, then there exists a unique $\hat{b}^A \in \bits$ and $\hat{x}^A \in \mx$ such that $f_{k^A, \hat{b}^A}(\hat{x}^A) = c^A$, so the post-measurement state is 
\begin{equation}
\ket{\psi^A_1} = \ket{\hat{b}} \ket{\hat{x}} \,. \label{eqn:post_g}
\end{equation}
The post-measurement states on Bob's side are analogous.

In \cref{step3}, the device receives a challenge type $\ct^A$ from Alice, and $\ct^B$ from Bob.
If $\ct^A = \texttt{a}$, the honest device simply measures the entire state $\ket{\psi^A_1}$ in the computational basis and returns the outcome $z^A$ to Alice.
The check applied by Alice is the following: let $z^A_1$ be the first bit of $z^A$, and $z^A_r$ the remainder of the string. Then, Alice checks that 
\begin{equation}
f_{k^A, z^A_1}(z^A_r) = c^A \,.
\end{equation}
From \cref{eqn:post_f} and \cref{eqn:post_g}, it is easy to see that this check passes for the honest device, irrespective of whether $k^A \in \kf$ or $k^A \in \kg$.
The honest behaviour and checks are analogous on Bob's side.

If $\ct^A =$\texttt{b}, the honest device measures the second register of the state $\ket{\psi^A_1}$ in the Hadamard basis and returns the outcome $d^A$ to Alice.
In the case where $k^A \in \kf$, this leaves the first register in the state (up to normalization)
\begin{equation}
\ket{\psi^A_2} = \ket{0} + (-1)^{d^A \cdot (x^A_0 \oplus x^A_1)} \ket{1} \,,
\end{equation}
where ``$\cdot$'' denotes the inner product between bit strings.
In the case where $k^A \in \kg$, this leaves the first register in the state 
\begin{equation}
\ket{\psi^A_2} = \ket{\hat{b}^A} \,.
\end{equation}
The analogous statement holds on Bob's side.

Alice and Bob now send questions $x, y \in \{\texttt{Computational}, \texttt{Hadamard}\}$ to their respective components of the device.
The honest device uses a pre-shared EPR pair to execute the teleportation circuit from the main text (\cref{eqn:teleport_circ}) on the state $\ket{\psi^A_2} \ket{\psi^B_2}$, obtaining $h^A, h^B \in \bits$, applies a Hadamard gate on the second qubit, and finally measures the two qubits in the basis given by the questions $x$ and $y$, respectively, obtaining $a, b \in \bits$ as outcomes.
The bits $h^A$ and $a$ are returned to Alice, and $h^B$ and $b$ to Bob.

We call the state right before the measurement $\ket{\psi_3}$.
Using \cref{eqn:teleportation_outcome} from the main text and commuting the Hadamard gate past the correction operator, we get (up to a global phase)
\begin{equation}
\ket{\psi_3} = (\sigma_X^{h^A} \sigma_Z^{h^B} \ot \sigma_X^{h^A} \sigma_Z^{h^B})(\1 \ot H ) CZ \ket{\psi^A_2}\ket{\psi^B_2} \,. \label{eqn:psi_3}
\end{equation}

To understand the checks applied by Alice and Bob, note that unless $k^A$ and $k^B$ are both in $\kf$ (i.e., $\theta^A = \theta^B = \texttt{Hadamard}$), the state $\ket{\psi_3}$ is still a product state (i.e., we are considering a product round).
Further, Alice and Bob know $h^A, h^B, d^A$, and $d^B$, and they can compute $x^A_0$ and $x^A_1$ or $\hat{x}^A$ from $c^A$ (and the same on Bob's side).\footnote{This computation requires a trapdoor, which is a piece of secret information that Alice and Bob generated when they produced the keys $k^A$ and $k^B$, but that is not shared with the device. Hence, the device cannot (efficiently) do this computation itself, which turns out to be crucial for the security of the protocol in \cite{comp_self_testing}.}
Hence, Alice and Bob know which product state the honest device has prepared, and they check whether the answers returned by a (potentially dishonest) device are the same as what the honest device would have returned.
Clearly, this means that an honest device succeeds with probability 1 in a product round.

In the case $\theta^A = \theta^B = \texttt{Hadamard}$ (called a Bell round in the main text), let us first consider the state without the correction operator:
\begin{equation}
\ket{\psi_3'} = (\1 \ot H) CZ \ket{\psi^A_2}\ket{\psi^B_2} \,.
\end{equation}
By a direct calculation, one can verify that (up to a global phase)
\begin{equation}
\ket{\psi_3'} = \ket{\phi^{(d^A \cdot (x^A_0 \oplus x^A_1), \; d^B \cdot (x^B_0 \oplus x^B_1))}} 
\end{equation}
where 
\begin{equation}
\ket{\phi^{(s^A, s^B)}} = (\sigma_Z^{s^A} \sigma_X^{s^B} \ot \1) \frac{\ket{00} + \ket{11}}{\sqrt{2}}
\end{equation}
are the 4 Bell states as in the main text.
It is easy to see that up to a global phase (which depends on $s^A, s^B, h^A$, and $h^B$, but which we can drop), the Bell states are invariant under the correction operator: 
\begin{equation}
(\sigma_X^{h^A} \sigma_Z^{h^B} \ot \sigma_X^{h^A} \sigma_Z^{h^B} )\ket{\phi^{(s^A, s^B)}} \equiv \ket{\phi^{(s^A, s^B)}} \,. 
\end{equation}
Therefore, we have (up to global phase)
\begin{align*}
\ket{\psi_3} 
&= (\sigma_X^{h^A} \sigma_Z^{h^B} \ot \sigma_X^{h^A} \sigma_Z^{h^B}) \ket{\psi_3'} \\ 
&= (\sigma_X^{h^A} \sigma_Z^{h^B} \ot \sigma_X^{h^A} \sigma_Z^{h^B}) \ket{\phi^{(d^A \cdot (x^A_0 \oplus x^A_1), \; d^B \cdot (x^B_0 \oplus x^B_1))}} \\
&\equiv \ket{\phi^{(d^A \cdot (x^A_0 \oplus x^A_1), \; d^B \cdot (x^B_0 \oplus x^B_1))}} \,. \numberthis \label{eqn:psi_3_bell}
\end{align*}

Note that as in the previous case, Alice and Bob can determine from the device's responses which Bell state $\ket{\phi^{(s^A, s^B)}}$ the honest device would have prepared:
Alice can compute $x^A_0$ and $x^A_1$ from $c^A$ (and similarly Bob computes $x^B_0$ and $x^B_1$ from $c^B$), and from \cref{eqn:psi_3_bell} we have: 
\begin{align*}
s^A &= d^A \cdot (x^A_0 \oplus x^A_1) \,, \\
s^B &= d^B \cdot (x^B_0 \oplus x^B_1) \,. \numberthis \label{eqn:compute_s}
\end{align*}

The Bell states are uniquely characterized as joint eigenstates of $\sigma_X \ot \sigma_X$ and $\sigma_Z \ot \sigma_Z$, where the eigenvalues depend on $s^A$ and $s^B$. For example, $\ket{\phi^{(0, 1)}}$ is the unique state that is a (+1)-eigenstate of $\sigma_X \ot \sigma_X$ and a (-1)-eigenstate of $\sigma_Z \ot \sigma_Z$.
For questions $x = y \in \{\texttt{Computational}, \texttt{Hadamard}\}$, both components of the honest device will measure the same Pauli observable and report back the results $a$ and $b$, respectively.
Hence, Alice and Bob, knowing $s^A$ and $s^B$, can check whether $a \oplus b$ is the correct eigenvalue.  

In summary, the winning condition in the Bell case is as follows:
\begin{itemize}[label={--}, noitemsep, topsep=0pt]
\item if $x = y = \texttt{Computational}$, the device wins if $a \oplus b = d^B \cdot (x^B_0 \oplus x^B_1)$.
\item if $x = y = \texttt{Hadamard}$, the device wins if $a \oplus b = d^A \cdot (x^A_0 \oplus x^A_1)$.
\item if $x \neq y$, the device always wins.
\end{itemize}
From \cref{eqn:psi_3_bell}, it is clear that the honest device always wins in a Bell round.

\bibliography{main}

\end{document}